\documentclass[usenatbib,twocolumn,useAMS]{mn2e}
\usepackage{graphicx}    
\usepackage{latexsym}
\usepackage{dcolumn}
\usepackage{bm}
\input{epsf}
\def\araa{Annu. Rev. Astron. Astrophys.} 
\def\prd{Phys. Rev. D} 
\def\apj{Astrophys. J.}         
\def\mnras{Mon. Not. R.  Astron. Soc.}
\def\aap{Astron. \& Astrophys.}
\newcommand{\bfig}{\noindent\begin{minipage}{3.48in}}
\newcommand{\efig}{\bigskip\end{minipage}}
\title{The effect of cluster magnetic field on the Sunyaev Zeldovich
power spectrum}
\author[Pengjie Zhang]
{Pengjie Zhang\thanks{E-mail:zhangpj@fnal.gov}\\
NASA/Fermilab Astrophysics Group,
Fermi National Accelerator Laboratory, Box 500,
Batavia, IL 60510-050}
\begin{document}      
\maketitle
\begin{abstract}
Precision  measurements of the Sunyaev Zeldovich (SZ) effect in
upcoming blank sky surveys
require theoretical understanding of all physical processes with  $\ga
10\%$ effects on the SZ power spectrum. We show that, observed cluster magnetic field could
reduce the SZ power spectrum by $\sim 20\%$ at $l\sim 4000$, where the
SZ power spectrum will be precisely measured 
by the Sunyaev Zeldovich array (SZA) and the Atacama cosmology
telescope (ACT).  At
smaller scale, this effect is larger and could reach a factor of
several. Such effect must be considered for an unbiased interpretation
of the SZ data. Though the magnetic effect on the SZ power spectrum is
very similar to that of radiative cooling, it is measurable by
multi-band CMB polarization measurement. 

\end{abstract}
\section{Introduction}
Our universe is almost completely ionized at $z\la 6$. Ionized
electron scatters off CMB photons by its thermal motion and so
generates secondary CMB temperature fluctuations. This effect is known
as the thermal  Sunyaev Zeldovich (SZ) effect. Since all free
electrons participate in the inverse Compton scattering and contribute
to the SZ effect, the SZ effect is an unbiased probe of the thermal
energy of the universe. Its precision measurement and interpretation
are of great importance to understand the thermal history of the universe. 

Several detections of the CMB excess power over the primary CMB at
$\sim 10^{'}$ scale (CBI at $l\la 3500$ \citep{Bond02,Mason03}, BIMA
at $l\sim 6800$ \citep{Dawson02} and a marginal detection by ACBAR at 
$l\sim 3000$ \citep{Kuo02}) may signal the first detections of the SZ effect in a
blank sky survey. Several upcoming CMB experiments such as ACT, Planck
and SZA
are likely able to measure the SZ power spectrum with $\sim 1\%$
accuracy\footnote{\citet{Zhangpj03} estimate the accuracy of the kinetic
SZ effect measurement by ACT, which could reach $1\%$. Since the
thermal SZ effect is about $30$ times stronger than the kinetic SZ
effect, a $1\%$ accuracy measurement of the thermal SZ effect by ACT
is highly likely.}. 

Such precision measurement of the SZ effect requires an accurate
understanding of it. Since upcoming SZ experiments are mainly
interferometers, most works have focused on the prediction of the SZ
power spectrum.  Lots effort has been made to predict
the SZ effect in an adiabatically evolving universe, both analytically
\citep{Cole88,Makino93,Atrio-Barandela99,Komatsu99,Cooray00,Molnar00, 
Majumdar01,Zhang01,Komatsu02} and simulationally
\citep{daSilva00,Refregier00,Seljak01,Springel01,Zhang02}. But various
processes could introduce $\ga 10\%$ uncertainties to the predicted power
spectrum. The most significant ones may be feedback, preheating and
radiative cooling, as favored by observations of  clusters
(\citet{Xue03} and  reference therein) and the soft X-ray background
\citep{Pen99}. The 
injection of non-gravitational energy heats up gas, makes it less clumpy and decreases
the SZ power spectrum \citep{daSilva01,Lin02,White02}. An energy injection of
$\sim 1 $KeV per nucleon could decrease the SZ power spectrum by a
factor of $2$.  Radiative cooling efficiently removes hot gas in the
core of clusters and groups 
and reduces the SZ power significantly,
especially at small scales \citep{daSilva01,Zhang03}.  
Supernova remnants generated by first stars cool mainly
through Compton scattering over CMB photons. The high efficiency of
such energy injection into CMB  could introduce a SZ effect
comparable to that 
of low redshift gas \citep{Oh03}. 

In this paper, we discuss the influence of cluster magnetic field on
the SZ effect. As we will find,  cluster magnetic field can suppress
the SZ power spectrum by $\sim 20\%$ at $l\sim 4000$ and a factor of
$2$ at $1^{'}$.

Micro-gauss magnetic field universally exists in intracluster medium
(ICM) (refer to \citet{Carilli02} for a recent review).  The strength
of magnetic field in the core of non-cooling flow clusters is
generally several $\mu$G, as  inferred from Faraday rotation measure
(\citet{Carilli02,Eilek02,Taylor02}, but see
\citet{Newman02,Rudnick03} for the discussion of smaller values).  The magnetic
pressure in the center of cluster could reach $\sim 1$-$10\%$ of the
gas thermal pressure. This extra pressure offsets part of the gravity
and prohibits ICM to further fall in and so results in  a less clumpy gas
core. The cluster SZ temperature decrement could be then reduced by a
factor $10\%$ \citep{Dolag00,Koch03}. However, this effect is
only non-trivial for low mass clusters and groups in which gravity is
weaker. Since such clusters are difficult to detect, the detection
of the magnetic field effect on individual clusters is highly challenging. 

The SZ power
spectrum, on the other hand, avoids this problem. Three
characteristics of the SZ power spectrum amplify the effect of
magnetic field comparing to that of individual clusters. (1) Less massive
clusters and groups are more populous, which amplifies their
contribution to the 
SZ effect. (2) The contribution of less massive clusters and groups to
the SZ power  spectrum concentrates on smaller angular scales,
comparing to more   massive clusters.  (3) At $l\ga
4000$, the contribution to the SZ effect is mainly from $z\ga 0.5$ 
\citep{Zhang01}, where less massive clusters are more dominant
comparing to nearby universe.
Combining these three points, the small scale SZ power is dominated by
low mass clusters and groups. Since
the influence of magnetic field on those clusters and groups is
larger,  one expects, if there is no strong decrease in the
strength of  magnetic field at $z\sim 1$, as suggested by high
redshift source rotation measures (\citet{Carilli02} and reference
therein), magnetic field could change the SZ power spectrum at small
scale by a significant fraction. Such effect is likely observable
in future SZ surveys, thus
the study of the effect of magnetic field on the SZ power spectrum
serves for both the precision modeling of the SZ effect and a better
understanding of cluster magnetic field and so deserves a detailed
analysis.

\citet{Koch03} build an analytical model to estimate the
effect of magnetic field on individual clusters adopting an
isothermal  $\beta$
model for the ICM state and solve the magneto-hydrostatic
equilibrium equation perturbatively.  Our goal is to investigate its
collective effect on the SZ power spectrum. For this purpose, we build a more
detailed and more consistent model, based on the model of {\it universal 
gas density profile} \citep{Komatsu01}. Since for 
low mass clusters and groups, magnetic
pressure is comparable to gas thermal pressure, the perturbative
method breaks down. So we solve the magneto-hydrostatic equilibrium equation 
non-perturbatively. We develop our model
in \S \ref{sec:model} and apply it to the SZ effect in \S
\ref{sec:SZ}. We discuss and conclude in \S
\ref{sec:discussion}. Throughout this paper, we 
adopt a WMAP-alone cosmology: $\Omega_m=0.268$,
$\Omega_{\Lambda}=0.732$, $\Omega_b=0.044$, $\sigma_8=0.84$ and
$h=0.71$ \citep{Spergel03}. 

\section{The effect of magnetic field on clusters}
\label{sec:model}
In this section, we will solve the hydrostatic equilibrium equation
of individual clusters for the gas density and temperature
profile, when magnetic field is present or not. First we need to know the
gravitational potential well, which is mainly determined by dominant
dark matter.  The dark matter  density 
profile can be well approximated by the NFW profile \citep{Navarro96,Navarro97}
\begin{equation}
\rho_{\rm dm}=\rho_s y_{\rm dm}(x)=\rho_s \frac{1}{x(1+x)^2}.
\end{equation}
Here, $x\equiv r/r_s$ is the radius in unit of core radius $r_s$. The
core radius is related to the virial radius by the compact factor
$c\equiv r_{\rm vir}/r_s$. According to the definition of virial
radius, $r_{\rm vir}\equiv [M/(4\pi \Delta_c(z)
\rho_c(z)/3)]^{1/3}$ where $\Delta_c\sim 100$ is the mean density of a
halo with mass $M$ inside of  its virial radius in unit of the critical density
$\rho_c(z)$ at redshift $z$. We adopt the predicted $\Delta_c$
from \citet{Eke96}. The compact factor we adopt is \citep{Seljak00}:
\begin{equation}
c=6\left(\frac{M}{10^{14}M_{\sun}/h}\right)^{-0.2}.
\end{equation}  

\begin{figure}
\epsfxsize=9cm
\epsffile{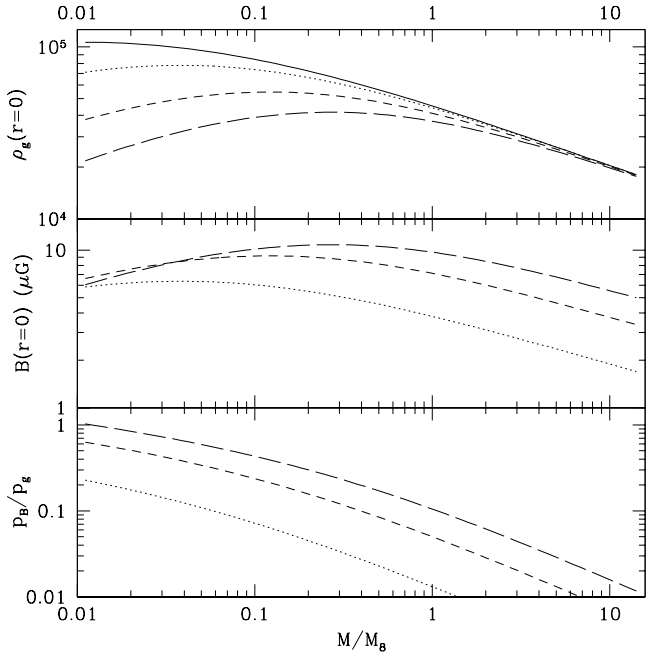}
\caption{The effect of magnetic field on intracluster gas state. We assume
$B(r)=B_*[\rho_g/10^4\bar{\rho}_g]^{0.9}$. The central gas density (in
unit of mean gas density),
strength of central magnetic field and the ratio between the magnetic
pressure and the thermal pressure are shown in top, middle and bottom
panel, respectively. For solid, 
dot, short dash and long dash lines, $B_*=0,1,2,3\ \mu G$,
respectively. More massive clusters have stronger 
gravity and larger gas pressure, so the effect of magnetic field is
weaker. For less massive clusters, magnetic gas pressure is comparable
with the thermal pressure and $\rho_g(r=0)$ can be greatly suppressed
by a factor of unity. This density suppression in turn suppresses the
strength of the magnetic field by the $B$-$\rho_g$ correlation.  So the
resulting strength of central magnetic field has only a weak dependence on halo
mass and falls in the observed range  of $1$-$10 \mu G$ over a broad
range of halo mass, from galaxy-size halos to most massive
clusters. On the other 
hand, magnetic field only weakly changes the gas temperature.\label{fig:B}}
\end{figure}

The gas density $\rho_g$ and
temperature $T_g$ can then be solved by the hydrostatic
equilibrium condition and one extra assumption on the gas state. In
the literature, one either 
assumes (a) that gas follows dark matter ($\rho_g\propto\rho_{\rm
dm}$. e.g. \citet{Wu02,Xue03}), or (b) isothermality ($T_g=$ cons.),
or (c) a polytropic gas  
($T_g\propto \rho_g^{\gamma-1}$. e.g. \citet{Komatsu01}). We will
follow the procedure of 
\citet{Komatsu01} and adopt the assumption (c). We will outline its
basic idea  in \S\ref{subsec:B=0} and extend it to 
the case when magnetic field exists (\S\ref{subsec:B}).

\subsection{$B=0$}
\label{subsec:B=0}
The three key ingredients of \citet{Komatsu01} are
\begin{itemize}
\item The hydrostatic equilibrium condition:
\begin{equation}
\label{eqn:hydroB=0}
\frac{dp_g}{dr}=-\frac{GM(\leq r)}{r^2}\rho_g.
\end{equation}

Here, $M(\leq r)$ is the total mass contained in the sphere with
radius $r$. It can be approximated as 
\begin{equation}
M(\leq r)\equiv4\pi \rho_s (1+\frac{\Omega_b}{\Omega_{\rm dm}}) r_s^3 m(x).
\end{equation}
Here, $m(x)\equiv\int_0^x u^2 y_{\rm dm}(u)du$. 
\item A polytropic form of the gas equation of state:
 \begin{equation}
p_g\propto \rho_g^{\gamma}. 
\end{equation}
Then Eq. (\ref{eqn:hydroB=0}) becomes
\begin{equation}
\label{eqn:ygB=0}
y_g^{\gamma-1}=1-3\frac{T_{\rm vir}}{T_g(r=0)} \frac{\gamma-1}{\gamma}
\frac{c}{m(c)} \int_0^x du \frac{m(u)}{u^2}.
\end{equation}
Here, $y_g(x)\equiv \rho_g(r)/\rho_g(r=0)$ is the relative gas density
profile. $T_{\rm vir}$ is the virial temperature 
defined as
\begin{equation}
T_{\rm vir}\equiv \frac{GM\mu m_H}{3k_Br_{\rm vir}}=5.23\Omega_m
\frac{M/M_8}{r_{\rm vir}/(\rm Mpc/h)} {\rm KeV}.
\end{equation}
Here $M_8=5.96\Omega_m M_{\sun}/h$ is the average mass contained
within a $8$ Mpc/h comoving radius. Thus the gas state is solved up to
three constants  
$\rho_g(r=0)$, $T_g(r=0)$ and $\gamma$.

\item Gas density distribution follows dark matter density
distribution in the outer  region of each cluster. So, $y_g(x)$ and
$y_{\rm dm}(x)$ must have the same slope at $x\ga c/2$. This
requirement simultaneously fixes $T_g(r=0)$ and $\gamma$:
\begin{eqnarray}
\label{eqn:T0B=0}
\frac{T_g(r=0)}{T_{\rm
vir}}&=&\frac{3}{\gamma}\frac{c}{m(c)}|S_*|^{-1}\\ \nonumber
&\times&\left(\frac{m(c)}{c}+(\gamma-1) |S_*|
\int_0^c du \frac{m(u)}{u^2}\right),
\end{eqnarray}
\begin{equation}
\label{eqn:gamma}
\gamma=1.15+0.01(c-6.5).
\end{equation}

Here, $S_*\equiv d\ln y_{\rm dm}(x)/\ln(x)|_{c}$ is the slope of the dark
matter density profile at $x=c$. $\rho_g(r=0)$ is
fixed by requiring that 
the baryon-dark 
matter density ratio in the outer region follows its universal ratio:
\begin{equation}
\rho_g(r=0)=\rho_s \frac{\Omega_b}{\Omega_{\rm dm}}\frac{y_g(c)}{y_{\rm dm}(c)}.
\end{equation}

\end{itemize}

\begin{figure}
\epsfxsize=9cm
\epsffile{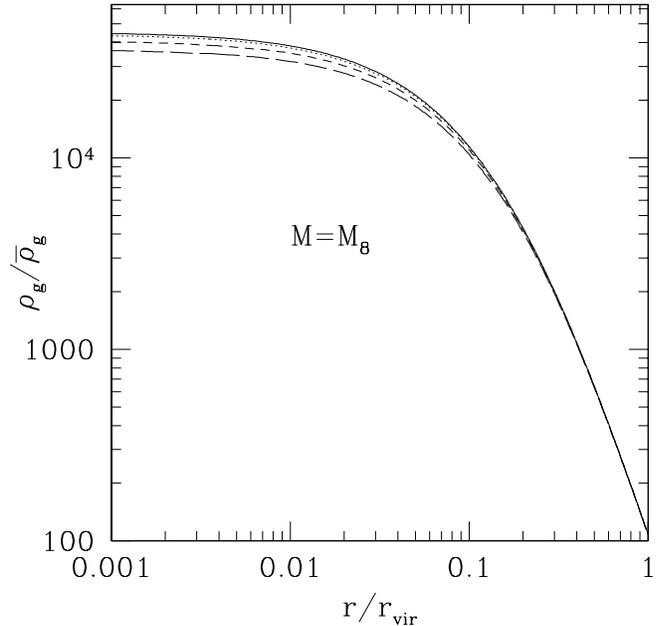}
\caption{The effect of magnetic file on the gas density profile.  We assume
$B(r)=B_*[\rho_g)/10^4\bar{\rho}_g]^{0.9}$. For solid,
dot, short dash and long dash lines, $B_*=0,1,2,3\ \mu G$
respectively. Since the magnetic field is 
strongest in the core, its effect is most significant in the core and
decreases dramatically outward. \label{fig:y}}
\end{figure}

\subsection{$B\neq 0$}
\label{subsec:B}

Magnetic pressure ($p_B=B^2/8\pi$) provides extra force to offset gravity and
suppresses the falling of gas into the gravitational potential
well. One then expects a less clumpy gas core. Clusters generally have
a magnetic field of the order $\mu G$, whose pressure is $\sim 10\%$  of
the thermal pressure: 
\begin{equation}
\frac{p_B}{p_g}=0.0252(\frac{1000}{1+\delta_g})(\frac{5 {\rm KeV}}{k_B
T_g}) (\frac{B}{1\mu  
G})^2 (\frac{0.02}{\Omega_b h^2}).
\end{equation}

With the existence of magnetic field, Eq. \ref{eqn:hydroB=0}
changes to 
\begin{equation}
\label{eqn:hydroB}
\frac{dp_g}{dr}+\frac{dp_B}{dr}=-\frac{GM(\leq r)}{r^2}\rho_g.
\end{equation}
If we assume that $B\propto \rho^\alpha$, we obtain
\begin{eqnarray}
y_g^{\gamma-1}+\eta y_g^{2\alpha-1}=(1+\eta)y_g^{\gamma-1}(B=0),
\end{eqnarray}
where 
\begin{equation}
\label{eqn:eta}
\eta=\frac{p_B(B,r=0)}{p_g(B,r=0)} \frac{2\alpha}{2\alpha-1}
\frac{\gamma-1}{\gamma}
\end{equation}
is determined by the central density and temperature.

There is no definite prediction of $\alpha$, but one can infer its lower
limit. 
Since most, if not all, magnetic field generation mechanisms, such as
galactic winds and hierarchical mergers of cluster formation,  produce
magnetic field positively correlated with gas density
(\citet{Carilli02} and reference therein), the further amplification of
adiabatic compression will produce $\alpha\geq 2/3$. For example,
the hierarchical merger produces $\alpha\simeq 0.9$
\citep{Dolag01}. In this paper, we consider two cases of $\alpha$,
$\alpha=0.9$ and $\alpha=2/3$.

\begin{figure}
\epsfxsize=9cm
\epsffile{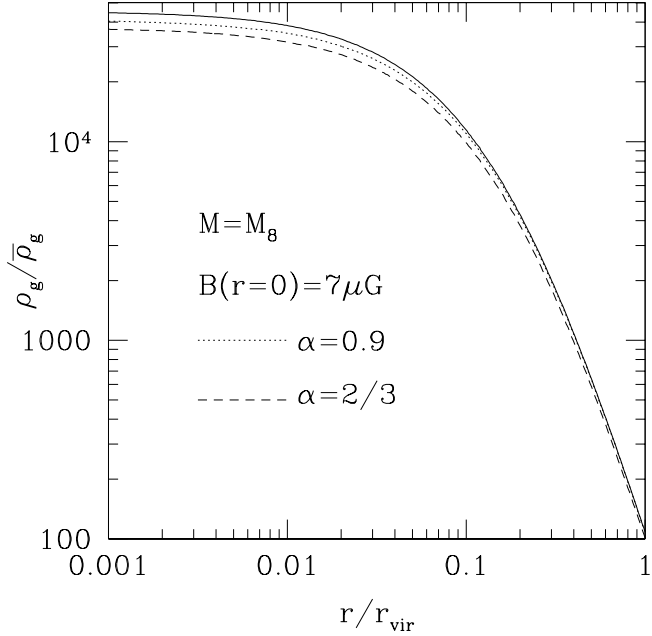}
\caption{The dependence of magnetic effect on individual clusters on the
$B$-$\rho_g$ correlation ($B\propto \rho_g^{\alpha}$). We choose suitable
coefficient in this 
scaling relation such that the strength of central magnetic field
($7\mu G$ in this figure) is identical for various
$\alpha$. Since the magnetic pressure of $\alpha$
drops more slowly than that of $\alpha=0.9$, its effect is observable
from the core to cluster outer region. So the overall effect of
magnetic field of $\alpha=2/3$ is larger than that of $\alpha=0.9$. \label{fig:dify}}
\end{figure}
 
Given  $\alpha\geq 2/3$, the magnetic pressure drops
faster than the thermal pressure, so in the outer regions near virial
radius, magnetic field 
can be neglected.  One can then easily show that $\gamma$ does not
change due to the presence of magnetic field. The two constant 
$T_g(r=0)$ and $\rho_g(r=0)$ are related to their corresponding values
when $B=0$ by:
\begin{equation}
\frac{T_g(r=0,B)}{T_g(r=0,B=0)}=(1+\eta)^{-1},
\end{equation}
and  
\begin{equation}
\frac{\rho_g(r=0,B)}{\rho_g(r=0,B=0)}=(1+\eta)^{\frac{-1}{\gamma-1}}.
\end{equation}

$\eta$ is fixed by 
\begin{equation}
\eta (1+\eta)^{(2\alpha-\gamma)/(\gamma-1)}=\eta_0,
\end{equation}
where $\eta_0$ is the value of $\eta$ by substituting  $\rho_g(r=0,B)$
and $T_g(r=0,B)$ in Eq. \ref{eqn:eta} with $\rho_g(r=0,B=0)$ and
$T_g(r=0,B=0)$. 

We adopt a parametric form of $B(r)$
\begin{equation}
\label{eqn:B}
B(r)=B_*\left(\frac{\rho_g(r)}{10^4\bar{\rho}_g(z=0)}\right)^{\alpha}.
\end{equation}
The results for various $B_*$ and $\alpha$ are shown in
Fig. \ref{fig:B}, \ref{fig:y} and \ref{fig:dify}. As
expected, magnetic field has weaker effect on more massive clusters
since their gravity is stronger.  For $M_8$ clusters,  the magnetic
pressure can reach $10\%$  of the thermal pressure and can suppress
the central gas density by $\sim 10\%$. 

Strong magnetic
field suppresses the gas infall (top panel, Fig. \ref{fig:B}) and
therefore the strength of magnetic 
field in turn by the $B$-$\rho_g$ correlation.   Such back-reaction is dominant in  low mass clusters and
groups and so causes the strength of magnetic field to cease to
increase toward low mass end (middle panel, Fig. \ref{fig:B}). Our parametric
model predicts a several $\mu G$ magnetic field over a wide range of
halo mass, from  galaxy-size halo mass 
($\la 0.01 M_8$) to massive cluster mass. We do not find any strong
dependence of magnetic field on halo mass,  which is consistent with
observations of galaxy (\citet{Beck96} and references therein) and
cluster magnetic field.  The agreement between our
predicted galaxy magnetic field strength with observations
suggests that our parametric treatment of magnetic field may extend to
galaxy scale. If so, magnetic field may suppress gas infall to low
mass halos by a factor of several. Such suppression   may have
a significant effect on star formation in such halos and may be
partly responsible for the formation of dark satellite halos due
to inefficient gas accretion.  This issue may deserve further
investigation.

We focus on a $M_8$ cluster to investigate the dependence of magnetic
effect on $B$-$\rho_g$ correlation. To
single out its effect,  we choose corresponding $B_*$ for $\alpha=0.9$
and $\alpha=2/3$ such that  the magnetic field in the core of a $M_8$
cluster is the  same for two cases. In order to satisfy this
requirement, $B_*$ for $\alpha=2/3$ must be larger than that of 
$\alpha=0.9$. So magnetic field of $\alpha=2/3$ has a larger pressure gradient
and thus a larger effect on clusters (Fig. \ref{fig:dify}). We have
adopted $B_*(\alpha=2/3)=2.94\ \mu G$ while $B_*(\alpha=0.9)=2.0\ \mu
G$ in Fig. \ref{fig:dify}. Since for $\alpha=2/3$, magnetic pressure drops
more slowly toward outer region than $\alpha=0.9$ case,  its effect
extends to a larger radius
(Fig. \ref{fig:dify}). So, its overall effect is significantly larger
than $\beta=0.9$ case. Since $\alpha$ is likely bigger than $2/3$,
this case should be treated as the upper limit of the magnetic effect.

\section{The Sunyaev Zeldovich effect}
\label{sec:SZ}

\begin{figure}
\epsfxsize=9cm
\epsffile{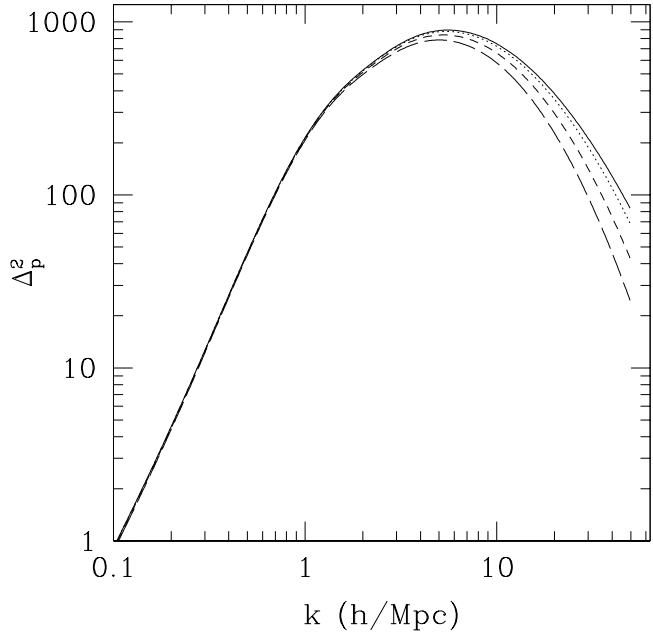}
\caption{The effect of magnetic field on the gas thermal pressure power
spectrum $\Delta^2_p$. $p\equiv (1+\delta_g) T_g/$KeV is the
dimensionless thermal pressure.  This definition differs from the usual
definition of the pressure power
spectrum by a factor $\bar{T}_g^2$.   We assume
$B(r)=B_*[(\rho_g/10^4\bar{\rho}_g(z=0)]^{0.9}$. For solid, 
dot, short dash and long dash lines, $B_*=0,1,2,3\ \mu G$,
respectively. The smaller scale power is mainly contributed 
by less massive clusters since they have smaller core radius. So, the
effect of magnetic field increases toward smaller scale. \label{fig:p}}
\end{figure}

\begin{figure}
\epsfxsize=9cm
\epsffile{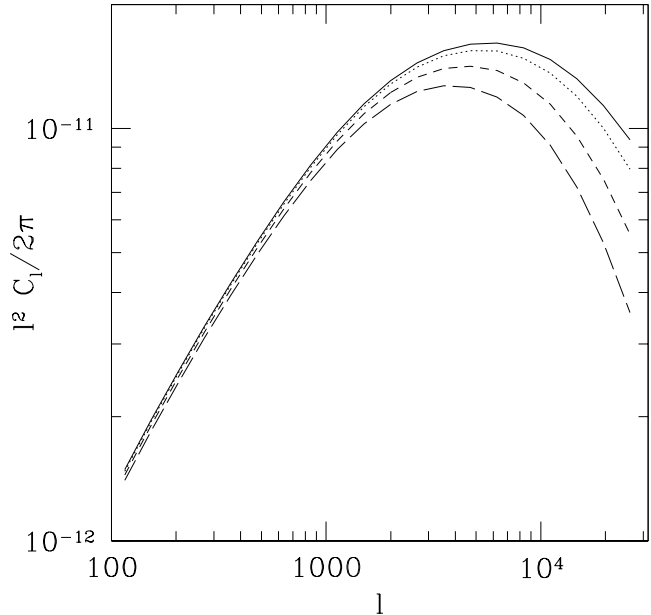}
\caption{The effect of magnetic field on the SZ effect power spectrum
$C_l$.  We assume
$B(r)=B_*[(\rho_g/10^4\bar{\rho}_g(z=0)]^{0.9}$.
For solid, dot, short dash and long dash lines, $B_*=0,1,2,3\ \mu G$,
respectively. Similar to the case of gas
pressure power spectrum, the effect of B increases toward small
scales where most contribution comes from less massive
clusters. \label{fig:cl}} 
\end{figure}

In the Rayleigh-Jeans regime, the SZ temperature decrement is given by
\citep{Zeldovich69}
\begin{eqnarray}
\frac{\Delta T}{T_{\rm CMB}}&=&-2\int\sigma_T \frac{n_e k_BT_g}{m_ec^2} ad\chi\\ \nonumber
&=&-2.37\times 10^{-4} \Omega_bh \int \frac{(1+\delta_g)k_BT}{\rm
keV}a^{-2} d\tilde{\chi}. 
\end{eqnarray}
Here, $\chi$ and $a$ are the comoving distance and scale factor,
respectively. $\tilde{\chi}\equiv \chi/(c/H_0)$ is the dimensionless
comoving distance while $H_0$ is the present Hubble constant. We
define a dimensionless gas pressure $p\equiv 
(1+\delta_g)k_BT_g/$KeV. Then the mean SZ decrement is determined by
the gas density weighted temperature $\bar{T}_g\equiv\langle
p\rangle$. In the halo model, each halo has a pressure distribution
$p(r,M)$, whose integral over the halo volume gives the contribution
of each halo to the mean SZ temperature decrement. Once one knows the
halo mass function $n(M,z)$, one can calculate
$\bar{T}_g$  by 
\begin{equation}
\label{eqn:meanT}
\bar{T}_g=\int \frac{dn}{dM} dM\left[\int_0^{r_{\rm vir}} p(r,M)a^{-3}
4\pi r^2dr\right].
\end{equation}
The halo mass function $n$ is well described by the Press-Schechter
formalism \citep{Press74}:
\begin{equation}
\frac{dn}{dM}=\sqrt{\frac{2}{\pi}}\frac{\rho_0}{M^2}\frac{\delta_c}{\sigma(M)}
\vert\frac{d\ln \sigma(M)}{d\ln M}\vert \exp [-\frac{\delta^2_c}{2\sigma^2(M)}].
\end{equation}
Here, $\rho_0$ is the present mean matter density of the
universe. $\sigma(M,z)$ is the linear theory rms density fluctuation
in a sphere containing mass $M$ at redshift $z$. $\delta_c$ is the
linearly extrapolated over-density at which an object virializes. For a
$\Omega_m=1$ universe, $\delta_c=1.686$. Since its dependence on
cosmology is quite weak \citep{Eke96}, we fix $\delta_c=1.686$. This
simplification introduces at most $1\%$ error.

The contribution of each halo to the SZ power spectrum is determined
by the Fourier component of its pressure profile:
\begin{equation}
\label{eqn:fourier}
p(k,M)=\int_0^{r_{\rm vir}} p(r,M) \frac{\sin(kr/a)}{kr/a} a^{-3}4\pi r^2dr.
\end{equation}
The collective contribution of all halos at a certain redshift is the pressure
power spectrum
\begin{eqnarray}
p^2(k)&=&\int p^2(k,M)\frac{dn}{dM} dM \\ \nonumber
&+&P_{\rm dm}(k)\left( \int p(k,M)b(M)\frac{dn}{dM}dM\right)^2.
\end{eqnarray}
Here, $b(M)$ is the linear bias of the halo number
overdensity with respect to dark matter overdensity 
such that the halo-halo power spectrum   $P(k,M_1,M_2)=P_{\rm
dm}(k)b(M_1)b(M_2)$.  The linear dark matter power spectrum $P_{\rm
dm}(k)$ is calculated by the BBKS transfer function fitting formula
\citep{BBKS}. We adopt the \citet{Mo96} formula to calculate 
$b(M)$. 

Throughout this paper, we alternatively refer the pressure variance
$\Delta^2_p\equiv p^2(k)k^3/2\pi$ as the pressure power
spectrum. $\Delta^2_p(k,z=0)$ for $\alpha=0.9$ is shown in
Fig. \ref{fig:p}. As expected, the effect of magnetic field
concentrates on small scales. At $k\sim 6 h/$Mpc, the peak of
$\Delta^2_p(k)$, it reduces $\Delta^2_p$ by $\sim 10\%$. At smaller
scales dominated by less massive halos, magnetic field can suppress
$\Delta^2_p$ by as large as a factor of $2$.

The Limber's integral of $\Delta^2_p$ over comoving distance  is the
2D SZ power  spectrum:
\begin{equation}
\frac{l^2C_l}{2\pi}=\pi\left(2.37\times 10^{-4} \Omega_bh\right)^2
\int \Delta^2_p(\frac{l}{\chi},z)a^{-4}
\frac{\tilde{\chi}}{l}d\tilde{\chi}. 
\end{equation}
\begin{figure}
\epsfxsize=9cm
\epsffile{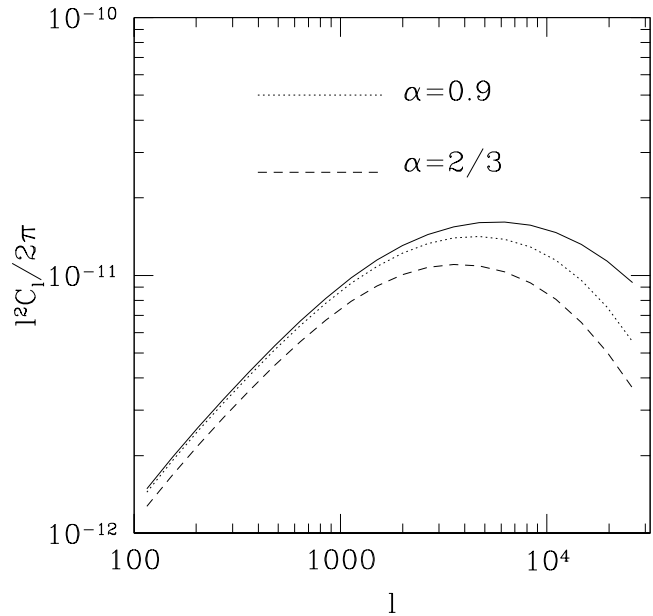}
\caption{The effect of $B$-$\rho_g$ correlation on the SZ power
spectrum. We have 
chosen $B_*=2.94\ \mu G$ for $\alpha=2/3$ such that the central
magnetic field of a $M_8$ cluster at $z=0$ is the same as that of the
case with $\alpha=0.9$ and $B_*=2.0\ \mu G$. Since magnetic pressure
drops more slowly  toward cluster outer region in the $\alpha=2/3$
case,  its effect extends in a wider spacial range and so becomes
observable even at angular scales as large as $l\sim 100$. Since
$\alpha$ is likely 
bigger than $2/3$, $\alpha=2/3$ case may show the upper limit of the
magnetic effect.  \label{fig:difcl}}
\end{figure}
\begin{figure}
\epsfxsize=9cm
\epsffile{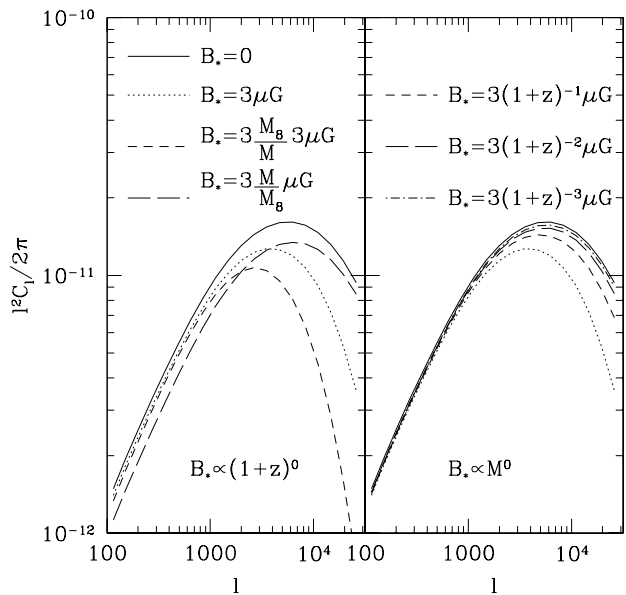}
\caption{The effect of the $B_*$ dependence on cluster mass and
redshift.  We parameterize this dependence  as a power law $B_*\propto
M^{a}(1+z)^b$. $C_l$ at larger 
scales is determined by more massive clusters. For $a>0$, more massive
clusters have stronger magnetic field, which suppresses $C_l$ at larger
scales. At smaller scales $C_l$ is determined by less massive
clusters and groups. For $a>0$, less massive clusters have weaker
magnetic field 
and the magnetic field effect on $C_l$ at small scales is suppressed. 
For $a<0$, the suppression on $C_l$ at small scales is much more
significant. If $B_*$ evolves faster than $(1+z)^{\sim -2}$, the
magnetic effect on $C_l$ is effectively negligible. \label{fig:difcl2}}
\end{figure}

To do this integral, one needs to know the evolution of cluster
magnetic field. The origin of cluster magnetic field
( \citet{Carilli02} and reference therein) is quite unclear, so in the
literature,  there is no consensus reached on the evolution of
magnetic field. For example,  the scenario of hierarchical merger of cluster
formation  predicts 
a very strong evolution $B(z)\propto10^{-2.5 
z}$ \citep{Dolag02}, while  magnetization mechanism associated with magnetized
galactic  winds generated by starbursts predicts a much slower
evolution \citep{Volk00}.  Given such divergent predictions, the most
reliable way to infer cluster magnetic field evolution may be
observations.  The measured large Faraday  rotation measures of 
high $z$ sources suggest that $\mu G$ magnetic field may have
existed in (proto-)cluster atmospheres (\citet{Carilli02} and
reference therein) at $z\ga 2$. Such magnetic field strength is
comparable to that of the present clusters. So, we assume
no evolution in the magnetic field and apply Eq.(\ref{eqn:B}) to all
$z$. 

We show the SZ power spectrum of $\alpha=0.9$ under such assumption in
Fig. \ref{fig:cl}. Since at higher redshifts, there are fewer massive
clusters, the effect of magnetic field is stronger on $C_l$ than on
$\Delta^2_p(z=0)$.  At $l\sim 4000$,
the peak of the SZ power spectrum, magnetic field can suppress the SZ power
spectrum by $\sim 20\%$. At large angular scale $l\la 1000$, where
nearby massive clusters dominate, the effect of magnetic field is
negligible. At $l\ga 4000$, the magnetic
suppression is significant  
and can reach a factor of several. Such effect must be taken into
account for a precision understanding of the SZ effect. 

For the choice\footnote{As a reminder, such choice of $B_*$ produces the same
central magnetic field for $M_8$ clusters at $z=0$} of  $(\alpha,B_*)=(0.9,2.0\mu {\rm G})$ and $(2/3,2.94 \mu {\rm
G})$, we
compare the resulted SZ power spectra. Since $\alpha=2/3$ suppresses
more on cluster gas density in a wider range (Fig. \ref{fig:dify}),
its effect on the 
SZ power spectrum is larger and extends to larger angular scales
(Fig. \ref{fig:difcl}). For the same reason, we find a  $30\%$
decrease in the  mean temperature decrement comparing to $B=0$ case
and a $20\%$ decrease comparing to $\alpha=0.9$ case. Since $\alpha$ is
likely bigger than  
$2/3$, $\alpha=2/3$ case should be treated as an upper limit of the
magnetic effect.

The above discussion has omitted any dependence of $B_*$ on
cluster mass and redshift. Though this special choice is consistent with
observations, since there is no solid predictions or measurements
on $B_*$, one has to be aware of other possibilities. In order to discuss the  effect of possible dependence of $B_*$ on $M$ and
$z$,  we parameterize 
this dependence as $B_*\propto M^a (1+z)^b$ and try several choices of
$a$ and $b$.  $a\ga 1$
produces too weak $B$ for small groups while $a\la -1$ 
produces too weak $B$ for massive clusters. So, we only discuss the cases of
$a=-1$ and $1$. For $b\la -3$, cluster magnetic field  at
$z>1$ is too weak and contradicts with observations, so, we only
discuss the cases  of $b=-3,-2,-1$. The result is shown in
fig. \ref{fig:difcl2}. If the main growth 
mechanism of cluster  magnetic field  is hierarchical merger, one then
expects $a>0$ since more massive clusters emerge from  more
mergers.  Such
correlation increases the magnetic field of more massive 
clusters and thus decreases $C_l$ at larger scales comparing to the $a=0$
case.  Cluster $B$ generation mechanism associated with star formation
tends to have $a<0$ since star forming galaxies mainly reside in field
(or equivalently, small halos)  instead of massive clusters. For
this case, we expect a larger suppression at small scales due to
stronger magnetic field in less massive clusters and groups. For the
redshift dependence, we find that if $B_*$ evolves faster than
$(1+z)^{\sim -2}$, the effect of magnetic field on $C_l$ is
negligible. Due to large uncertainties in both $a$ and $b$, we
postpone further discussion in this paper. 

\section{Discussion}
\label{sec:discussion}
Cluster magnetic field affects not only the SZ effect. Here we address
its influence on cluster entropy,  X-ray luminosity and the soft X-ray
background (XRB).

Magnetic field may be partly responsible for entropy floors observed in
clusters (e.g. \citet{Xue03} and reference therein). A
conventional definition of cluster entropy is 
$K=T_g\rho_g^{-2/3}$. Magnetic field does not change entropies of
very massive clusters, but it increases entropies of less massive
clusters by decreasing their gas densities. But for a reasonable magnetic
field, such effect is not sufficient to explain the entire entropy floors
observed and could only be a minor cause comparing to preheating,
feedback and radiative cooling \citep{Xue03}. 

The X-ray luminosity of clusters have a strong dependence on gas
density ($\propto \rho_g^2$) and mainly comes from central regions of
clusters. So one expects a larger suppression to cluster
X-ray luminosity than to cluster SZ effect. Since less massive halos
are more clumpy, their contribution to the soft XRB is larger. So the
suppression of magnetic field to individual cluster X-ray luminosity
is further amplified in the soft XRB. One then expects a order of
unity suppression to the mean flux and power spectrum of the soft XRB
contributed by clusters and groups.  This effect can be quantitatively
investigated following a 
similar procedure, as applied to the XRB in the literature
\citep{Wu03,Zhangpj03a}.

Besides its dynamic effect, magnetic field can further change the ICM
state by the suppression of the  thermal conduction (\citet{Malyshkin01} and 
reference therein). This  could produce
non-negligible effect on the Sunyaev Zeldovich effect, especially when
preheating, feedback or radiative cooling are present.  This effect
does not change the mean SZ temperature decrement, but would likely
increase the small scale SZ power by hot and cool patches caused by
inefficient heat conduction. Since this issue 
requires a detailed understanding of the distribution of entangled magnetic
field in a cluster, which is not available at present, we postpone such
estimation. 

Here we address several  subtleties in our SZ calculation. One is how to deal
with the gas loss caused by magnetic field. Since magnetic field
suppresses the gas infall and we only integrate over the virialized
volume (see Eq. \ref{eqn:meanT} and \ref{eqn:fourier}), the gas mass
in such integration of each cluster is less than that of $B=0$
case. One can always  increase the integral upper limit in
Eq. \ref{eqn:meanT} and 
\ref{eqn:fourier} to compensate the gas mass loss. Unfortunately, the fate
of such gas is unclear. But because gas outside
of virial radius 
is several times cooler than gas in the core, such effect is minor, so
we neglect such calculation in this paper. Another issue is
$\gamma$. We have assumed a constant $\gamma$ across 
each clusters, but the effect of magnetic field on the gas polytropic
state is unclear. Though these issues have to be scrutinized in
magneto-hydrodynamic (MHD)
simulations, the consistency between our analytical predictions and
MHD simulations \citep{Dolag00} suggests that our model should be
appropriate at the  first order approximation. 

We then conclude that, cluster magnetic field suppresses the SZ power
spectrum around its peak 
by $\sim 20\%$ and by a factor of $\sim 2$ at $\sim 1^{'}$. Such
effect must be considered for an accurate theoretical understanding
of the SZ effect and an unbiased interpretation of the SZ measurement in future
blank sky surveys. The effect of 
magnetic field to the SZ power spectrum is similar to that of
radiatively cooling \citep{Zhang03}, which also decreases the SZ
power spectrum at small scales significantly while leaves the large
scale power spectrum barely touched. Such similarity brings extra
difficulty to 
extract the magnetic effect from the SZ observation and therefore the
interpretation of the SZ data.  The CMB
polarization measurement can be applied to recover cluster magnetic fields
\citep{Ohno03} and helps for a robust prediction of the SZ effect
under the presence of magnetic field.

{{\bf \it Acknowledgments}}: I thank Scott Dodelson, Lam Hui and Bing
Zhang for helpful discussions. This work was supported by the DOE and the
NASA grant NAG 5-10842 at Fermilab.


\begin{thebibliography}{}
\bibitem[Atrio-Barandela \& Mucket(1999)]{Atrio-Barandela99}
Atrio-Barandela, F.; Mücket, J. P., 1999, \apj, 515, 465
\bibitem[Bardeen et al.(1986)]{BBKS} Bardeen, J. M.; Bond, J. R.;
Kaiser, N.; Szalay, A. S., 1986, \apj, 304, 15
\bibitem[Beck et al.(1996)]{Beck96} Beck, Rainer; Brandenburg, Alex;
Moss, David; Shukurov, Anvar; Sokoloff, Dmitry; 1996, \araa, 1996, 34, 155
\bibitem[Bond et al.(2002)]{Bond02} Bond, J.R.;  Contaldi, C.R.;
Pen, U.L.;  Pogosyan, D.;  Prunet, S.;  Ruetalo, M.I.;  Wadsley, J.W.;
Zhang, P.J.;  Mason, B.S.;  et al., 2002, submitted
to \apj, astro-ph/0205386
\bibitem[Carilli \& Taylor(2002)]{Carilli02} Carilli, C.L.; Taylor,
G.B., 2002, \araa, 40, 319
\bibitem[Cole \& Kaiser(1988)]{Cole88} Cole, Shaun; Kaiser, Nick,
1988, \mnras, 233, 637
\bibitem[Cooray, Hu \& Tegmark(2000)]{Cooray00} Cooray, Asantha; Hu,
Wayne; Tegmark, Max; 2000, \apj, 540, 1
\bibitem[da Silva et al.(2000)]{daSilva00} da Silva, Antonio C.;
Barbosa, Domingos; Liddle, Andrew R.; Thomas, Peter A., 2000, \mnras,
317, 37
\bibitem[da Silva et al.(2001)]{daSilva01} da Silva, Antonio C.; Kay,
Scott T.; Liddle, Andrew R.; Thomas, Peter A.; Pearce, Frazer R.;
Barbosa, Domingos; 2001, \apj, 561, 15L
\bibitem[Dawson et al.(2002)]{Dawson02} Dawson, K. S.; Holzapfel,
W. L.; Carlstrom, J. E.; Joy, M.; LaRoque, S. J.; Miller, A. D.;
Nagai, D; 2002, \apj, 581, 86
\bibitem[Dolag \& Schindler(2000)]{Dolag00} Dolag, K.; Schindler, S.,
2000, \aap, 364, 491
\bibitem[Dolag et al.(2001)]{Dolag01} Dolag, K.; Schindler, S.;
Govoni, F.; Feretti, L., 2001, \aap, 378, 777
\bibitem[Dolag, Bartelmann \& Lesch(2002)]{Dolag02} Dolag, K.;
Bartelmann, M.; Lesch, H., 2002, \aap, 387, 395
\bibitem[Eilek \& Owen(2002)]{Eilek02} Eilek, J.; Owen, F.N., 2002,
\apj, 567, 202
\bibitem[Eke, Cole \& Frenk(1996)]{Eke96} Eke, Vincent R.; Cole,
Shaun; Frenk, Carlos S., 1996, \mnras, 282, 263
\bibitem[Koch, Jetzer \& Puy(2003)]{Koch03} Koch, P.M.; Jetzer, Ph.;
Puy, D., 2003, New Astronomy, 8, 1
\bibitem[Komatsu \& Kitayama(1999)]{Komatsu99} Komatsu, Eiichiro;
Kitayama, Tetsu, 1999, \apj, 526, L1
\bibitem[Komatsu \& Seljak(2001)]{Komatsu01} Komatsu, E. \&
Seljak, U., 2001, \mnras, 327, 1353
\bibitem[Komatsu \& Seljak(2002)]{Komatsu02} Komatsu, E.; Seljak, U., 2002,
\mnras, 336, 1256
\bibitem[Kuo et al. (2002)]{Kuo02} Kuo, C.L.;  Ade, P.A.R.;
Bock, J.J.; Cantalupo, C.; Daub, M.D.; Goldstein, J.;
Holzapfel, W.L.; Lange,A.E.; Lueker,M.; et al., 2002,
submitted to \apj, astro-ph/0212289
\bibitem[Lin et al.(2002)]{Lin02} Lin,Kai-Yang; Lin, Lihwai;
Woo, Tak-Pong; Tseng, Yao-Hua; Chiueh, Tzihong; 2002, submitted to ApJL,
astro-ph/0210323
\bibitem[Majumdar(2001)]{Majumdar01} Majumdar, Subhabrata, 2001, \apj,
555, L7
\bibitem[Makino \& Suto(1993)]{Makino93} Makino, N.; Suto,
Y., 1993, \apj, 405, 1
\bibitem[Malyshkin(2001)]{Malyshkin01} Malyshkin, L., 2001, \apj, 554,
561
\bibitem[Mason et al.(2003)]{Mason03} Mason, B. S.; Pearson, T. J.;
Readhead, A. C. S.; Shepherd, M. C.; Sievers, J.; Udomprasert, P. S.;
Cartwright, J. K.; Farmer, A. J.; Padin, S.; Myers, S. T.; et al.,
2003, \apj, 591, 540
\bibitem[Mo \& White(1996)]{Mo96} Mo, H. J.; White, S. D. M., 1996,
\mnras, 282, 347
\bibitem[Molnar \& Birkinshaw(2000)]{Molnar00} Molnar, S. M.;
Birkinshaw, M., 2000, \apj, 537, 542
\bibitem[Navarro, Frenk \& White(1996)]{Navarro96} Navarro, Julio F.;
Frenk, Carlos S.; White, Simon D. M., 1996, \apj, 462, 563
\bibitem[Navarro, Frenk \& White(1997)]{Navarro97} Navarro, Julio F.;
Frenk, Carlos S.; White, Simon D. M., 1997, \apj, 490, 493
\bibitem[Newman, Newman \& Rephaeli(2002)]{Newman02} Newman, W.; Newman, A.;
Rephaeli, Y., 2002, \apj, 575, 755
\bibitem[Oh, Cooray \& Kamionkowski(2003)]{Oh03} Oh, S.Peng; Cooray,
A. \& Kamionkowski, M., 2003, submitted to \mnras, astro-ph/0303007
\bibitem[Ohno et al.(2003)]{Ohno03} Ohno, Hiroshi; Takada, Masahiro;
Dolag, Klaus; Bartelmann, Matthias; Sugiyama, Naoshi; 2003, \apj, 584,
599
\bibitem[Pen(1999)]{Pen99} Pen, Ue-Li, 1999, \apj, 510, L1
\bibitem[Press \& Schechter(1974)]{Press74}  Press, William H.;
Schechter, Paul, 1974, \apj, 187, 425
\bibitem[Refregier et al.(2000)]{Refregier00} Refregier, Alexandre;
Komatsu, Eiichiro; Spergel, David N.; Pen, Ue-Li, 2000, \prd, 6113001
\bibitem[Rudnick \& Blundell(2003)]{Rudnick03} Rudnick, Lawrence;
Blundell, Katherine M., 2003, \apj, 588,143
\bibitem[Seljak(2000)]{Seljak00} Seljak,U., 2000, \mnras, 318, 203
\bibitem[Seljak, Burwell \& Pen(2001)]{Seljak01} Seljak, Uros;
Burwell, Juan; Pen, Ue-Li, 2001, \prd, 63, 063001
\bibitem[Spergel et al.(2003)]{Spergel03} Spergel, D.N.; Verge, L.;
Peiris, H.V.; Komatsu, E.; Nolta, M.R.; Bennett, C.L.; Halpern, M.;
Hinshaw, G.; Jarosik, N.l; et al. 2003,
astro-ph/0302209 
\bibitem[Springel, White \& Hernquist(2001)]{Springel01} Springel,
Volker; White, Martin; Hernquist, Lars, 2001, \apj, 549, 681
\bibitem[Taylor, Fabian \& Allen(2002)]{Taylor02} Taylor, G.B.;
Fabian, A.C.; Allen, S.W., 2002, \mnras, 334, 769
\bibitem[Volk \& Atoyan(2000)]{Volk00} Volk, H.J.; Atoyan, A.M., 2000,
\apj, 541, 88
\bibitem[White, Hernquist \& Spingel(2002)]{White02} White, Martin;
Hernquist, Lars; Springel, Volker, 2002, \apj, 577, 569L
\bibitem[Wu \& Xue (2002)]{Wu02} Wu, Xiang-Ping; Xue, Yan-Jie; 2002,
\apj, 572, 19
\bibitem[Wu \& Xue(2003)]{Wu03} Wu, Xiang-Ping; Xue, Yan-Jie, 2003,
\apj, 590, 8
\bibitem[Xue \& Wu(2003)]{Xue03} Xue, Yan-Jie; Wu, Xiang-Ping, 2003,
\apj, 584, 34
\bibitem[Zeldovich \& Sunyaev(1969)]{Zeldovich69} Zeldovich, Y.B.;
Sunyaev, R., 1969, Ap\&SS, 4, 301
\bibitem[Zhang \& Pen(2001)]{Zhang01} Zhang, Pengjie; Pen, Ue-Li, 2001,
\apj, 549, 18
\bibitem[Zhang, Pen \& Wang(2002)]{Zhang02} Zhang, Pengjie; Pen,Ue-Li;
Wang, Benjamin, 2002, \apj, 577, 555
\bibitem[Zhang \& Pen(2003)]{Zhangpj03a} Zhang, Pengjie \& Pen, Ue-Li,
2003, \apj, 588, 704
\bibitem[Zhang, Pen \& Trac(2003)]{Zhangpj03} Zhang, Pengjie; Pen,
Ue-Li; Trac, Hy; 2003, astro-ph/0304534, accepted by  \mnras.
\bibitem[Zhang \& Wu(2003)]{Zhang03} Zhang, Yu-Ying; Wu, Xiang-Ping,
2003, \apj, 583, 529

\end{thebibliography}
\end{document}